This note explores the mathematical theory to solve modern gambler's ruin problems.  We establish a ruin framework and solve for the probability of bankruptcy.  We express the probability of bankruptcy and distance parameters needed.  And we also show how this relates to other dimensions of the problem from expected time to ruin and alteration from symmetrical up and down changes in trading performance.

In the 18th century, Bernoulli solved the original gambler's ruin problem.  This problem's solution reflects that a gambler with a fair chance or an equal gain or loss per wager, will eventually run out of gambling chips when playing against an opponent with infinite resources (e.g., chips).  Prior to Bernoulli's contribution, additional variations of the gambler's ruin problem were emerging throughout the 17$^{th}$ century, from famous European quantitative scientists such as Fermat, Pascal, and Huygens.  Such special enhancement to the problem included limited and distinct quantities of wagering chips between two opponents, or unequal chances of winning any particular trial.  In the past century, probabilists under a different premise known as the cliff hanger problem have reincarnated this problem[i].  Where a drunken man, who is initially a certain number of steps from a cliff, then proceeds to take randomly steps, either towards or away from the cliff.  Applications away from drunkards, gamblers, or both, do exist.  Particularly in the realm of biological reactions and the pathways of self growth and decay.

So in this note we apply this problem to a realistic options trading strategy where a modern gambler, instead of gaining or losing a fixed quantity of chips, can gain or lose a balanced exponential rate on their chips.  For example, a hypothetical strategy that can provide a 100% return if it was successful or loses 50% if it had failed.  It turns out that this complication of the gambler's ruin problem can be reduced to the original ruin problem if one calibrates the absorption barrier that defines bankruptcy.

We can connect this idea most readily to present-date option payoffs related onto the changes in interest rates. And in an rough, low-frequency example, we know that in the summer of 2011 the 10-year U.S. treasury note yielded 2.8%. So we suggest that in a small theoretical model that a year later we would have a 10-year yield that was either 5.6%, or 1.4%. This fits our 100% or 50% change pattern, respectively. And we know in the summer of 2012, with the 10-year yield of the same instrument at 1.4%, we could have again modeled a year ahead yield. This time at either 2.8%, or 0.7%. And again we know that by this summer of 2013, the 10-year yield was 2.8%.

That was a similar empirical template that mirrors the theoretical gambling framework we are solving for in this note. So instead of bankruptcy equaling 0 chips, we can establish a ruin threshold that is a portion of the original chips. Immediately below we will show this to be defined as the "distance" equaling:

**loss level = (½)$^{distance}$**
**log$_{½}$(loss level) = distance**

**Noting that we can reformulate distance as log(loss level)/log(½).**
**Or ln(loss level)/ln(½).**

For example, if we wish to establish a portfolio blow-up level of 25% of the original chips (e.g., 75% loss in portfolio), then the distance would be 2 trials. The first trial could halve the portfolio to 50% of the original, and the second trial could further halve this 50% so that we are at the level of 25% of the original. For this gambler's ruin problem we can use complete probability combinatorics and only need to examine a limited combinations of gains and losses. These gains or losses are at the end, just before bankruptcy. We will rationalize this below.

First, the ultimate wager can not be a gain and still result in a bankruptcy; the final trial must be a loss. Second, the penultimate wager must also be a loss. Certainly one can imagine this trial being a gain instead. However, in order for a "bankruptcy" to occur, ultimately any rebound at the penultimate wager towards the end has to eventually be a loss, in order to provide the final wager. Therefore in summary, one only needs to keep track of the final two trial combinations.

The combinatorics therefore can be shown below, separating out the initial trials when the numbers of gains are less than two trails. And we combine this with the probabilities associated with the remaining trails, which allow for the number of gains in total to be equal to or exceed two trials. The reasoning we have is that in cases when the total number of gains equal or exceeds two trials, then the number of losses is equal to the reversal of that number of gains, plus the distance number of trials. So when the gains equal two or more, then those could be the final two trials, and there is a possibility for all of the initial trials then equaling the distance number of losses. This component of the combinatorial probability is something we'll subtract out later in the proof below.

First let's show how the series looks, and in this example we are distance agnostic so that we can theoretically solve the information needed. Here are the variable definitions we will use:

*p*            = *probability of gain*
*q*            = *probability of loss*
             = *1-p*
*d*            = *distance*
*N*            = *number of gains total*

**Individual trial combination probability**      = $q^{(d+N)} * p^N$.

And before solving the combinations, we will quickly review the basics

of multinomial expressions (see note here for quadnomials, or intermediate level multinomials):

$$(n\ k_1, k_2, k_m) = n! / [k_1!\ k_2!\ \ldots\ k_m!]$$

*Where n is the total number of wagering trials, and $k_m$ represents the variables that are repetitive and so the permutations do not distinguish among one another in the ordering.*

So now combinations where the number of gain(s) is less than the distance, we solve for the specific multinomial needed for our specific gambler's ruin formulae:

*$(d + N*2 - 2)! / [(d+N-2)!\ N!]$ ,               if N<2*

*Noting that $(d+N-2)+N=(d+N*2-2)$.*

And in the remaining cases where the number of gains is greater than the distance, we remove those specific invalid probability combinations as we discussed further above:

*$(d + N*2 - 2)!/[(d+N-2)!\ N!] – (N*2 - 2)!/[(N-2)!\ N!]$,       if N$\geq$2*

*Noting that $(N*2 -2)=(N-2+N)$.*

So let's examine the initial probability series for the lowest numbers of total gains.

$q^{(d+0)} * p^0 (d + 0*2 - 2)! / [(d+0-2)!\ 0!]$
+
$q^{(d+1)} * p^1 (d + 1*2 - 2)! / [(d+1-2)!\ 1!]$
+
$q^{(d+2)} * p^2 [(d + 2*2 - 2)! / [(d+2-2)!\ 2!] – (2*2 - 2)!/[(2-2)!\ 2!]]$
+
$q^{(d+3)} * p^3 [(d + 3*2 - 2)! / [(d+3-2)!\ 3!] – (3*2 - 2)!/[(3-2)!\ 3!]]$
+
…

**Which simplifies to:**
$q^d$
+
$q^{(d+1)} * p * d$
+
$q^{(d+2)} * p^2 [(d+2)!/[d!*2]]$
+
$q^{(d+3)} * p^3 [(d+4)!/[(d+1)!\ 3!] – 4]$
+
…

**Or an expansive arithmetic-geometric series that approximates:**
$q^d(1 + q*p*d + q^2*p^2*[(d+2)*(d+1)]/2 + q^3*p^3*[(d+4)*(d+3)*(d+2)]/6 + …)$
$\approx q^d * [1/(1-q*p*d)]$
$\approx q^d * [1/(1-(q-q^2)*d)]$
$\approx q^d * [1/(1-q*d)]$,                              if $q \leq \frac{1}{2}$
and d is large
$\approx q^d / [1/p^d]$

This is not the method generally shown in advanced probability and statistics books, though it provides a clearer breakout using the series techniques familiar to advanced finance professionals.  This solution to our gambler's ruin problem shows the familiar outcome that there is a small non-bankruptcy probability, when the chance of a gain (i.e., **p**) is greater than 50%.  And the solution shows an eventual probability of bankruptcy otherwise.

Now there are two additional wrinkles that we can examine with this advanced gambler's ruin problem.  The first is to compute the expected time for bankruptcy within this gambler ruin framework.  And the second is how to deal with derivative strategies that do not have an exponential up and down move of 100% and -50%, respectively.  While the latter shows up periodically in probability literature[ii], the former generally does not.

To solve the first problem we begin by noting that the expected time can be solved as a probability summation of time until ruin, multiplied by the probabilities we have above for the same term of the series.  So the probability of 0 gains times the combinations of 0 gains, plus the probability of 1 gains times the combinations of 1 gains, plus probability of 2 gains times the combinations of 2 gains, etc.

Also at this very initial limit, if there are no gains, then the distance itself (i.e., **d**) is the smallest expected time until a gambler goes bankrupt.  And for each successive total gain, an additional two trials are necessary: one for the gain, and an additional one to reverse that gain prior to ruin.  We know that the theoretical ratio we have between the two trial probabilities in the series above is approximately $p^d$.  So the expected time to ruin would equal:

*[1/(1-$p^d$)]\*(d-1)+(d-1)*

**Do recall that d is <u>preferably</u> very large, so we do not need to worry about cases such as d=1.**

Now for the latter wrinkle to our ruin problem, let's explore the financial implications for scenarios where a trader's strategy does not involve up and down moves that are exponentially symmetrical.  As with binomial approximations to options pricing strategy, we can easily solve this by adjusting the gain and loss probabilities instead[iii],[iv].

Let's try this out with an example of the transformation.  Suppose that we have an initial 50-50 chance of exponentially gaining 75% or losing 75%, but now we want to adjust our gambler's ruin problem to consider a loss of 25% instead of a 75% loss.  No change to the gaining assumption of 75%.  Then this would be the modified probabilities to align the problem[v]:

*p\*(75%) + q\*(-75%)*
*= 50%\*(75%) + 50%\*(-75%)*
*= 0*
*= q'\*(75%) + p'\*(-25%)*
*= (1-p')\*(75%) + p'\*(-25%)*
*= p'\*(-100%) + 75%*

*So p' would equal 75/100, or 75%, instead of the original p of 50%.*

This would mean that we would neutralize our original 50-50 chance of being +75%/-75% to being a 75-25 chance of being +75%/-75%.  And thus, this later expression would equate to proportions of changing our 50-50 chance to a +75%/-25% performance.  Since each trial would artificially amplify the results, if one were to consider a focus on the distance to bankruptcy aspect of this problem, then they should want to keep the actual level of the modified gains and losses in mind.  This is so that they can ensure that they reasonably still work, particularly if **d** is not exceptionally large, or if the new loss assumption expands so that the probability of gains per wager suddenly falls below 50%.

In summary, this note explores the mathematical theory to solve modern gambler's ruin problems.  We establish a ruin framework and solve for the probability of bankruptcy.  We also show how this relates to the expected time to bankruptcy and review the risk neutral probabilities associated an adjustment to asymmetrical views.